\documentclass[12pt]{iopart}
\usepackage[english]{babel}
\usepackage{amssymb}
\usepackage{epsfig}
\begin{document}
\newcommand{\Od}{{\cal O}}
%\newcommand{\lsim}   {\mathrel{\mathop{\kern 0pt \rlap  {\raise.2ex\hbox{$<$}}}
%  \lower.9ex\hbox{\kern-.190em $\sim$}}}
%\newcommand{\gsim}   {\mathrel{\mathop{\kern 0pt \rlap
%  {\raise.2ex\hbox{$>$}}}
%  \lower.9ex\hbox{\kern-.190em $\sim$}}}

%\preprint{FTPI-MINN-08/09}
%\preprint{UMN-TH-2640/08}

\title{Is the CMB Cold Spot a gate to extra dimensions?}%

\author{J.A.R. Cembranos$^1$
, A. de la Cruz-Dombriz$^2$
, A. Dobado$^2$
 \,\,and A.L. Maroto$^2$
}
\address{
$^1$William I. Fine Theoretical Physics Institute,
University of Minnesota, Minneapolis, 55455, USA\\
$^2$Departamento de  F\'{\i}sica Te\'orica I,
Universidad Complutense de Madrid, 28040 Madrid, Spain.}
\date{\today}% It is always \today, today,

\begin{abstract}
One of the most striking features found in the cosmic
microwave background data
is the presence of an anomalous Cold Spot (CS) in the temperature
maps made by the Wilkinson Microwave
Anisotropy Probe (WMAP). This CS has been
interpreted as the
result of the presence of a collapsing texture, perhaps coming from
some early universe Grand Unified Theory (GUT) phase transition.
In this work we propose an alternative explanation based on
a completely different kind of texture which appears in a
natural way in a broad class of brane-world models.
This type of textures known as brane-skyrmions can be understood as
holes in the brane which  make  possible to
pass through them along the extra-dimensional space.
The typical scales needed for the proposed brane-skyrmions
 to correctly describe the observed CS can be as low as
the electroweak scale.

\end{abstract}

%\pacs{98.80.-k, 04.50.-h}% PACS, the Physics and Astronomy
\maketitle 

One of the most important pieces of information about the
history and nature of our universe comes from the Cosmic Microwave
Background (CMB). Measurements of the CMB temperature anisotropies
obtained by WMAP \cite{WMAP,Hinshaw} have been thoroughly studied in
recent years. Such anisotropies have been found to be Gaussian as
expected in many standard cosmological scenarios corresponding to
density fluctuations of one part in a hundred thousand in the early
universe. However, by means of a wavelet
 analysis, an anomalous CS, apparently inconsistent with homogeneous
Gaussian fluctuations, was found in \cite{Vielva,Cruz1} centered at
the position $b=-57$$^\circ$, $l=209$$^\circ$ in galactic
coordinates, with a characteristic scale of about 10$^\circ$. The
existence of this CS has been confirmed more recently in
\cite{Cruz2,Nas}.

What is the origin of this CS? A very interesting possibility has
been proposed recently in \cite{cruzturok}. According to it some
theories of high energy physics predict the formation of various
types of topological defects, including cosmic textures
\cite{turok1} which would generate hot and cold spots in the CMB
\cite{turok2}. These textures would be the remnants of a symmetry breaking
phase transition that took place in the early universe. In order to
produce textures, the cosmic phase transition must be related to a
global symmetry breaking pattern from one group $G$ to a subgroup
$H$ so that the coset space  $K=G/H$ has a non trivial third homotopy group. 
A  typical example is $K=SU(N)$, which has associated 
$\pi_3(K)= \Bbb{Z}$ for $N\geq 2$. 
Notice that, as usual, in order to get a texture formed
in the transition, the symmetry breaking must correspond to a global
symmetry since if it were local it could be gauged away.

Textures can be understood as localized wrapped field
configurations which collapse and unwind on progressively larger
scales. Textures can produce a concentration of energy which give
rise to a time dependent gravitational potential. CMB photons
traversing the texture region will suffer a red or a blue shift
producing a cold or hot spot in the CMB maps.
In \cite{cruzturok} the authors
consider a $SU(2)$ Non Linear Sigma Model (NLSM) to build up a
model of texture that could explain the observed CS at $b=-57$$^\circ$,
$l=209$$^\circ$. They simulate the unwinding texture by using a
spherically symmetric scaling solution and they find a fractional
temperature distortion given by:
\begin{equation}
\frac{\Delta
T}{T}(\theta)=\pm\epsilon\frac{1}{\sqrt{1+4(\theta/\theta_c)^2}}
\end{equation}
where $\theta$ is the angle from the center, $\epsilon$ is a measure
of the amplitude and $\theta_c$ is the scale parameter that depends
on the time at which the texture unwinds. The best fit of the CS is found 
for $\epsilon=7.7\times10^{-5}$ and $\theta_c=5.1$$^\circ$. In the above 
solution, this value for $\theta_c$ implies that the observed texture 
unwound at
$z\sim 6$. On the other hand the parameter $\epsilon$ 
is given by $\epsilon\equiv 8\pi^2 G \Phi_0$, 
where $\Phi_0$ is the fundamental symmetry breaking scale which is then
set to be $\Phi_0\simeq 8.7 \times 10^{15}$ GeV. This scale is
nicely close to the GUT scale thus making the results given in
\cite{cruzturok}  extremely interesting.

Nevertheless it is important to stress that textures require having
a global symmetry breaking but what one finds typically in GUTs is a
local symmetry breaking producing the Higgs mechanism and then
destroying the topological meaning of texture or any other possible
defect appearing in the cosmic transition. In this work we explore a
different possibility to obtain textures that could explain the CS.
This possibility is based on the so called brane world models which have
the advantage of providing global symmetries by assuming proper boundary 
conditions and symmetries for the fields propagating in the bulk space. 
In these models, our universe is considered to be a three dimensional brane 
with tension $\tau\equiv f^4$ living in higher $D=4+N$ dimensional space
(the bulk space) with the $N$ extra dimensions compactified to some
manifold $B$ with small volume $V_N$ \cite{ADD}. Then it is possible
to find the relation $M_P^2=V_NM_D^{2+N}$, where $M_P$ is the Planck
scale and $M_D$ is the fundamental gravity scale in $D$ dimensions
which, in the phenomenologically interesting cases, could be of the
order of the TeV (the electroweak scale). In this kind of
scenario, the Standard Model (SM) particles are not allowed to leave
the brane, but gravitons propagate through the whole bulk space. For
simplicity we will assume that the whole $D$ dimensional space can
be factorized as ${\it M}_D= {\it M}_4\times B$ with the brane lying
on the ${\it M}_4$ space-time manifold (for simplicity we will
assume $B$ to be homogeneous). Then we can choose the obvious
coordinates $\{x^\mu,y^m\}$ with $\mu=0,1,2,3$;  $m=1,2,...,N$ and
the ansatz for the bulk metric $G_{MN}=\mbox{diag}(\tilde
g_{\mu\nu}(x),\tilde g'_{mn}(y))$. In absence of the brane,  
this metric possesses an isometry group that is assumed to
be of the form $G({\it M}_D)=G({\it M}_4)\times G(B)$. The presence
of the brane spontaneously breaks the symmetry to some subgroup
$G({\it M}_4)\times H$ with $H\subset G(B)$ being some subgroup of $G(B)$. 
The position of the brane can be parametrized as
$\{x^\mu,Y^m(x)\}$ and its induced metric in the ground state is
just $g_{\mu\nu}\equiv \tilde g_{\mu\nu}\equiv G_{\mu\nu}$. However,
when brane excitations (branons) are present, the induced metric is
given by $g_{\mu\nu}=\tilde g_{\mu\nu}-\partial_\mu Y^m \partial_\nu
Y^n \tilde g'_{mn} $. On the other hand it is easy to show that $B$
is diffeomorphic to the coset $K=G(B)/H$ which is the space of the
Goldstone bosons (GB) associated to the spontaneous symmetry
breaking of the isometries -transverse translation- produced by the
presence of the brane. Thus the transverse translations of the brane
-branons- can be considered as GB on the coset $K$ and the branon
fields can be defined as coordinates  $\pi^\alpha $ on $K$ as:
$\pi^\alpha=(v/R_B)\delta_m^\alpha y^m$, where $R_B$ is the typical
size of the compactified space and $v\equiv f^2R_B$ is the typical
size of the coset $K$. Then, the induced metric on the brane can be
written in terms of the branon fields $\pi$ as:
\begin{eqnarray}
 g_{\mu\nu}=\tilde
g_{\mu\nu}-\frac{1}{f^4}\partial_\mu\pi^\alpha
\partial_\nu\pi^\beta h_{\alpha\beta}
\label{induced}
\end{eqnarray}
where $h_{\alpha\beta}$ is the $K$ metric which can be obtained
easily from the $B$ metric \cite{DM}. In more complicated cases, the
metric $g_{\mu\nu}$ could also be a function of the extra dimension
coordinates $y$.  Then it is possible to show that branons may become
massive. In fact in \cite{DM2} these massive branons were shown to
behave as weakly interacting massive particles (WIMPs) and thus being
natural candidates for dark matter in this kind of scenario.

However in this work we would like to study
another important property of the branon fields, namely the
possibility of wrapping around the extra dimension space $B$ giving
rise to non-trivial topological configurations as it was studied in
detail in \cite{CDM}. In particular, for $\pi_3(B)=\pi_3(K)=\Bbb{Z}$ we
may have texture-like configurations called brane-skyrmions. As we
will see these brane-skyrmions have some common features with the
textures considered in \cite{cruzturok} and therefore could provide
an explanation of the CS observed in the CMB in these kinds of extra
dimensions brane-world models.

Brane-skyrmions can be nicely
understood in geometrical terms as some kind of holes in the
brane that make it possible to pass through them along the $B$ space.
This is because in the core of the topological defect the symmetry
is restablished. In particular, in the case considered in this work
the broken symmetry
is basically the translational symmetry along the
extra-dimensions.

At low energies, brane dynamics can be described by the Nambu-Goto
action:
\begin{eqnarray}
S\,=\,-f^4\int \mbox{d}^4x\sqrt{-g}
\end{eqnarray}
with $g_{\mu\nu}$ being the induced metric defined in
(\ref{induced}). In order to simplify the calculations we will
consider the simplest case $B\simeq K \simeq SU(2) \simeq S^3 $ and
introduce spherical coordinates on both spaces, $M_4$ and $K$. In
$M_4$ we denote the coordinates $\{t,r,\theta,\varphi\}$ with $\phi
\in [0,2\pi)$, $\theta \in[0,\pi]$ and $r \in [0,\infty)$. On the
coset manifold $K$, the spherical coordinates are denoted
$\{\chi_K,\theta_K,\phi_K\}$ with $\phi_K \in [0,2\pi)$, $\theta_K
\in [0,\pi]$ and $\chi_K \in [0,\pi]$. Notice that
such coordinates cover the whole spherical manifolds and relate
to the physical branon fields (local normal geodesic coordinates on
$K$) by:
\begin{eqnarray}
\pi_1&=&v \sin\chi_K\sin\theta_K\cos\phi_K,\nonumber\\
\pi_2&=&v \sin\chi_K\sin\theta_K\sin\phi_K,\label{esfk}\\ \pi_3&=&v
\sin\chi_K\cos\theta_K.\nonumber
\end{eqnarray}

The coset metric in spherical coordinates is written as
\begin{eqnarray}
h_{\alpha\beta}=\left(
\begin{array}{ccc}
v^2&\\ &v^2\sin^2(\chi_K) &\\ & & v^2
\sin^2(\chi_K)\sin^2(\theta_K)
\end{array}\right).
\end{eqnarray}

In terms of branon fields, it can be shown that the Nambu-Goto action
can be expressed as a derivative expansion, whose lowest-order term
with two field derivatives is nothing but the $SU(2)$ NLSM.  
In spherical coordinates, the brane-skyrmion with winding number
$n_W$ is given by the non-trivial mapping
$\pi^\alpha:S^3\longrightarrow S^3$ defined from:
\begin{eqnarray}
\phi_K=\phi\,,\,\,\,\,\,\,\,
\theta_K=\theta\,,\,\,\,\,\,\,\,
\chi_K=F(t,r)\,,\label{skyan}
\end{eqnarray}
with boundary conditions satisfying $F(t,\infty)-F(t,0)=n_W\pi$. This
map is usually referred to as the hedgehog ansatz.

From the Nambu-Goto action, we obtain the equation of  motion for our
skyrmion profile $F(t,r)$:
%\begin{widetext}
\begin{eqnarray}
\mbox{sin}(2F)&-&2rF'+\Big(r^2+\frac{v^2}{f^4}\mbox{sin}^2 F\Big)\label{EqF}\\
&\times&\frac{\ddot{F}-F''+\frac{v^2}{f^4}(\ddot{F}F'^2
+F''\dot{F}^2-2F'\dot{F}\dot{F}')}{1-\frac{v^2}{f^4}(\dot{F}^2-F'^2)}\,=\,0
\nonumber 
\end{eqnarray}
%\end{widetext}
where dot and prime denote $t$ and
$r$ derivatives respectively.

In this work we are interested in the potential cosmological effects
due to the presence of a brane-skyrmion within our Hubble radius.
For that purpose we will
compute the gravitational field perturbations at large distances
compared to the the size of the extra dimensions, i.e., we are
interested in the $r^2\gg R_B^2=v^2/f^4$ region.
Notice that in this
region, gravity behaves  essentially as in four dimensional space-time and
standard  General Relativity can be used in the
calculations. Notice also that in order to simplify the calculations
we will ignore the effects due to the universe expansion.
This is justified provided $r\ll H_0^{-1}$. In such a case
the unperturbed (ignoring the defect presence) background metric can be
taken as Minkowski, i.e.,
$\tilde g_{\mu\nu}=\eta_{\mu\nu}$, and the above equation of motion reduces to:
\begin{eqnarray}
r^2(\ddot{F_0}-F_0'')+\mbox{sin}(2F_0)-2rF_0'\,=\,0
\label{EqFzeroth}
\end{eqnarray}
what is equivalent to expression $(3)$ in \cite{turok2}. 
Notice that this is an expected result since, as shown in \cite{turok1,Press},  
at large distances, i.e. except in the microscopic unwinding regions, 
the dynamical evolution of the fields are completely 
independent of the symmetry breaking mechanism, it simply depends on the 
geometry of the coset manifold $K$. On small scales \cite{turok1,Press} it is
possible that higher-derivative terms could affect the dynamics and even 
stabilize the textures, this is also the case of 
brane-skyrmions \cite{CDM}, although generically they could unwind by means of
quantum-mechanical effects. 

Eq. (\ref{EqFzeroth}) admits an exact solution with $n_W=1$:
\begin{eqnarray}
F_{0}(t,r)\,=\,2\,\mbox{arctan}(-r/t)
\label{F_0_exacta}
\end{eqnarray}
with $t<0$ since, in order to get a cold spot, photons should pass the
texture position before collapse.

Our approximated equation (\ref{EqFzeroth})
is consistent with (\ref{EqF}) for the
above solution since the  second term in the numerator
vanishes for  $F\equiv F_{0}$ and
\begin{eqnarray}
\dot{F}_{0}^2-F_{0}'^2\,&=&\,4(r^2-t^2)/(r^2+t^2)^2\nonumber \\
\sin^2 F_{0}\,&=&\,4r^2t^2/(r^2+t^2)^2
\end{eqnarray}
so that in the considered regime, the neglected terms are
irrelevant  for all $r$  and $t$ values.

Once $F(t,r)$ has been determined,
we may calculate the energy-momentum tensor components also in this
region from the Nambu-Goto action as: $T^{\mu\nu}= - 2\vert
\tilde{g} \vert^{-1/2}\delta S/\delta \tilde{g}_{\mu\nu}$.
In spherical coordinates they become
\begin{eqnarray}
T_{00}\,&=&\,\frac{2v^2(r^2+3t^2)}{(t^2+r^2)^2};\,\,\,\,
T_{rr}\,=\,\frac{2v^2(r^2-t^2)}{(t^2+r^2)^2}\nonumber\\
T_{0r}\,&=&\,-\frac{4v^2 r t}{(t^2+r^2)^2};
\,\,\,T_{\theta\theta}\,=\,\frac{2v^2 r^2(r^2-t^2)}{(t^2+r^2)^2}\nonumber\\
T_{\phi\phi}\,&=&\,\mbox{sin}^2 \theta \, T_{\theta\theta}
\label{T_spherical}
\end{eqnarray}
 Note that $\nabla_{\mu} T^{\mu}_{\nu}$  identically vanishes.
%%%%%%%%%%%%%%%%%%%%%%%%%%%%%%%%%%%%%%%%%%%%%%%%%%%%%%%%%%%%%%%%%%%%%%%%%%%%%%%%%%%%%%%%%%%%%%
We will now determine the background metric $\tilde g_{\mu\nu}$ in the
$r\gg R_B$ region and in  the presence of the  brane-skyrmion
as a small perturbation on the Minkowski metric, i.e.
$\tilde g_{\mu\nu}=\eta_{\mu\nu}+h_{\mu\nu}$.

Thus, for the scalar perturbation of the  Minkowski space-time
in the longitudinal
gauge we have:
\begin{eqnarray}
\mbox{d}s^2\,=\,(1+2\Phi)\mbox{d}\eta^2 -(1-2\Psi)
\delta_{ij}\mbox{d}x^{i}\mbox{d}x^{j}
\label{perturbed_metric}
\end{eqnarray}
where potentials $\Phi\equiv\Phi(t,\overrightarrow{x})$ and
$\Psi\equiv\Psi(t,\overrightarrow{x})$ and perturbed Einstein's
tensor components in cartesian coordinates are the following
(see \cite{Giovannini})
 \begin{eqnarray}
\delta\mathcal{G}^{0}_{0}\,&=&\,2\nabla^2\Psi\nonumber\\
\delta\mathcal{G}^{j}_{i}\,&=&\,[-2\ddot{\Psi}-
\nabla^2(\Phi-\Psi)]\delta_{i}^{j}
+\partial_{i}\partial^{j}[\Phi-\Psi]\nonumber\\
\delta\mathcal{G}^{0}_{i}\,&=&\,2\partial_{i}\dot{\Psi}
\end{eqnarray}
with $i,j=1,2,3$ and $\nabla^2\equiv\sum_{i=1}^{3}
\partial_{i}\partial^{i}$. Using Einstein equation
$\delta\mathcal{G}^{\mu}_{\nu}=8\pi\mbox{G} T^{\mu}_{\nu}$
we determine that perturbed quantities $\Phi$ and $\Psi$ are
\begin{eqnarray}
\Psi\,\equiv\,\Phi\,=\,4\pi\mbox{G}v^2\,\mbox{log}\frac{r^2+t^2}{t^2}\,.
\label{Phi}
\end{eqnarray}
The physical metric on which photons propagate is not
the $\tilde g_{\mu\nu}$ we have just calculated, but the
induced metric (\ref{induced}). However using the solution in
(\ref{F_0_exacta}), we  find that the contribution from branons
fields is $\Od( R_B^2/r^2)$, i.e. negligible, so that
 $g_{\mu\nu}\simeq \tilde g_{\mu\nu}$.

Photons propagating on the perturbed metric will suffer  red(blue)-shift
due to the Sachs-Wolfe effect. The full expression for the
temperature fluctuation is given by:
%\begin{widetext}
\begin{eqnarray}
\Big(\frac{\Delta T}{T}\Big)_{SW}\,
&=&\,- [\Phi]^{\tau_{f}}_{\tau_{i}}\,
+\,\int^{\tau_{f}}_{\tau_{i}}(\dot{\Psi}
+\dot{\Phi})\mbox{d}\tau\,\nonumber \\
&=&\,
- [\Phi]^{\tau_{f}}_{\tau_{i}}\,
+\,\int^{\tau_{f}}_{\tau_{i}}2\dot{\Phi}\mbox{d}\tau
\label{SW_effect}
\end{eqnarray}
%\end{widetext}
where we have considered local and integrated SW effects and
neglect Doppler contribution. $\tau_{i}$ is the  decoupling time
and $\tau_{f}$  the present time.
Subsitituting expression (\ref{Phi}) in the previous one
and using that $r^2\,=\,z^2+R^2$, and $z\,=\,t-t_{0}$, we get
\begin{eqnarray}
\Big(\frac{\Delta T}{T}\Big)_{SW}
\,&=&\,8\pi\mbox{G} v^2\Big[\frac{t_0}{\sqrt{2R^2+t_0^2}}\,
\mbox{arctan}\frac{t_0+2z}{\sqrt{2R^2+t_0^2}}\nonumber\\
&-&\mbox{log}|t_0+z|\Big]
\end{eqnarray}
where $R$ is the impact parameter and $t_0$ is the time at which the
photon passes the texture position at $z=0$.
In the limit where $z_{i}\rightarrow -\infty$ and
$z_f\rightarrow\infty$ the result is
\begin{eqnarray}
\Big(\frac{\Delta T}{T}\Big)_{SW}
\,=\,\epsilon\frac{t_0}{\sqrt{2 R^2+t_{0}^2}}
\end{eqnarray}
with $\epsilon\,\equiv\,8\pi^2\mbox{G}v^2$, in complete agreement with \cite{turok2}.

Let us estimate the scales required  for this kind of texture
to explain the observed CS.
For the minimal model supporting brane-skyrmions with three extra
dimensions $N=3$ we will have  $M_{P}^2\simeq R_B^3M_{D}^5$
with $M_{P}$ Planck mass, $M_{D}$ the fundamental scale of
gravitation in $D=7$  dimensions and $R_{B}^3$ the characteristic volume
for the compactified extra dimensions. Therefore
\begin{eqnarray}
R_{B}\,\simeq\,\Big(\frac{M_{P}^2}{M_{D}^5}\Big)^{1/3}
\end{eqnarray}
and consequently
\begin{eqnarray}
v^2\,\equiv\,f^4R_{B}^2\,\simeq\,f^4\Big(\frac{M_{P}^2}{M_{D}^5}\Big)^{2/3}\,.
\end{eqnarray}
In order to get $\epsilon\simeq 7.7 \cdot 10^{-5}$, we
need $v\simeq 1.2 \cdot 10^{16}$ GeV, which in fact can be
achieved with $M_D\sim f\sim$ TeV. Notice that for this parameter
range, the radius of the extra dimension is around $R_B\sim 10^{-8}$ m, i.e.
our approximations are totally justified. In addition, we have also checked 
that the possible effects coming from a non-vanishing branon mass
are suppressed by $\Od(M^2/v^2)$. Therefore for mass values also
around $M\sim $ TeV (which are typical of branon as dark 
matter, see \cite{DM2}) such effects are negligible. 
In other words,
brane skyrmions provide an accurate description for the CS without
the need of introducing very high energy (GUT) scales since
the correct temperature fluctuation amplitude can be
obtained with natural values around the electroweak scale.

Let us emphasize that there is no difference between the scenario 
where branons behave like
WIMPs and the one containing brane-skyrmions, except that the
latter is a subset of the former, since not all the brane-worlds support
textures. In other words, the same model can provide at the same
time WIMPs in the 
form of branons and
brane-skyrmions able to explain the cold spot. This is quite
remarkable, since it allows to explain both phenomena with the
same scale, which may be as low as the electro-weak scale.

In order to estimate the abundance 
of  brane-skyrmions in this model, it is important to take into account that 
their low-energy (large distance) dynamics  
is described by the same NLSM as that in \cite{cruzturok} and therefore, 
 the expected evolution
of textures in an expanding universe in the model presented in this work 
will be completely equivalent to that in the mentioned reference. This is 
nothing but the well-known
fact that except in the microscopic unwinding region, the field  evolution 
only depends on the geometry of the coset space $K$, but not on 
the details of the
symmetry breaking mechanism. As shown in \cite{Press} in a 
simple model with a 
potential term, the final abundance of defects and other properties of 
the pattern of density perturbations are expected to be not very  
sensitive to the short distance physics, once the texture unwinds, making
this kind of theories highly predictive. For that reason, we expect that
provided the same kind of initial conditions are imposed in both models,
the predicted abundance of hot and cold spots agree with that obtained
from simulations in \cite{cruzturok}. Such simulations show 
that the number of textures unwindings 
per comoving
volume and conformal time $\eta$ can be
estimated as $(dn/d\eta)=\nu\eta^{-4}$ with
$\nu\simeq 2$. This allows to estimate the number of hot and cold spots in
a given angular radius interval. As commented above, this is  
a quite robust result at late times with little effects form short
distance dynamics. 
In the case of brane skyrmions, the short distance
effects will be embodied in the higher-derivative terms appearing
in the expansion of the Nambu-Goto action or even in possible induced 
curvature terms generated by quantum effects \cite{CDM}.  
Provided that such terms do not stabilize the brane-skyrmions,  we expect 
that such an abundance  could be directly applied 
in our case.

Future experiments will be able to confirm the studied model. On 
the one hand, the fact that the fundamental scales of the theory are 
of the order of the TeV opens the possibility 
to test this explanation with collider experiments through the production
of real or virtual branons and KK-gravitons. 
%In the simplest model of three equal 
%extra dimensions compactified in  $S^3$, $f$ is typically one order of 
%magnitude bigger than ${M}_D$ if both scales are around the TeV scale. 
The expected signatures of the model \cite{Mirabelli:1998rt}  
from the 
production of KK-gravitons come fundamentally from the 
single photon channel studied by LEP, which restrict ${M}_D> 1.2$ TeV 
at 95\% of C.L. 
On the other hand, 
the LHC will be able to test the model up to ${M}_D=3.7$ TeV, analysing 
single photon and monojet production.

Finally,  there exists also the possibility to find 
signatures of the model at 
low energies associated with the branon (brane fluctuations) 
phenomenology. This case  is more interesting from the cosmological point of 
view, since branons can constitute the non-baryonic dark matter abundance as 
typical WIMPs \cite{DM2,Cembranos:2004jp}. Present  
constraints coming from the single photon analysis realized by L3 (LEP) 
imply $f > 122$ GeV (at 95 \% C.L.) \cite{Alcaraz:2002iu,Achard:2004uu} and 
the LHC will be able to check this model up to $f=1080$ GeV through 
monojet production
\cite{Cembranos:2004jp}. The idea to test the physics associated with the 
cold spot with the next generation of colliders at the TeV scale is a very 
intriguing and distinctive property of this texture.

%On the other hand,
%the cold spot will not be correlated with CMB polarization if it is 
%produced by a time-dependent metric as (\ref{perturbed_metric}), in contrast, 
%for example, with a primordial density fluctuation. We may also detect 
%the gravitational lensing angle of order $\epsilon\sim10^{-4}$ radians 
%created by the compact texture at $z\sim 6$. In addition, the texture should produce
%tens of smaller cold and hot spots \cite{cruzturok}. In particular, the blue-shifted photons 
%may give information of the particular symmetry breaking theory beyond the 
%low energy NLSM which leads to (\ref{EqFzeroth}). These photons are affected 
%by the texture after collapsing, and this process could provide a unique opportunity 
%to observe extra dimensions beyond standard perturbative analyses.

%\vspace{.1cm}

%\vspace{.1cm}
 {\bf Acknowledgements:}  This work
 has been partially supported by the DGICYT (Spain) under the
 projects FPA2004-02602, FPA2005-02327, CAM/UCM 910309, by
 UCM-Santander PR34/07-15875 and by DOE grant DOE/DE-FG02-94ER40823.
%\bibliography{DM12}% Produces the bibliography via BibTeX.

\end{document}